\documentclass[journal=jacsat,manuscript=article]{achemso}

\usepackage[version=3]{mhchem} 

\usepackage[utf8]{inputenc} 
\usepackage[T1]{fontenc}    
\usepackage{hyperref}       
\usepackage{url}            
\usepackage{booktabs}       
\usepackage{amsfonts}       
\usepackage{nicefrac}       
\usepackage{microtype}      
\usepackage{lipsum}
\usepackage{graphicx}
\graphicspath{ {./images/} }
\usepackage{amssymb}
\usepackage{amsmath}

\usepackage{booktabs}
\usepackage{siunitx}
\author{Vasileios Fotopoulos}
\email{vasileios.fotis.19@ucl.ac.uk}
\affiliation[University College London]
{Department of Physics and Astronomy, University College London, Gower Street, London WC1E 6BT, United Kingdom}
  \author{Alexander L. Shluger}
\affiliation[University College London]
{Department of Physics and Astronomy, University College London, Gower Street, London WC1E 6BT, United Kingdom}

\title{Atomistic Simulations of H–Cu Vacancy Cosegregation and H Diffusion in Cu Grain Boundary}

\begin{document}
\maketitle




\begin{abstract}
Hydrogen embrittlement remains a critical challenge in structural and electronic applications of copper (Cu) but its mechanism is still not fully understood. In this study, we combine density functional theory (DFT) and bond-order potential (BOP) simulations to determine the atomistic pathways for hydrogen adsorption/incorporation and fast interfacial diffusion at Cu grain boundaries (GBs), including its interaction with vacancies. Undercoordinated regions, such as surfaces and GBs, serve as preferential adsorption/incorporation sites for atomic hydrogen, especially in the presence of Cu vacancies. The presence of hydrogen in GB further enhances the segregation of Cu vacancies, leading to the formation of stable H--V$_{Cu}$ complexes with cosegregation energy gains of up to $-$0.8 eV. Furthermore, our simulations reveal that the migration barriers for hydrogen within the GB networks are as low as 0.2 eV and significantly lower than in bulk Cu (0.42~eV). The results presented in this paper suggest an atomistic mechanism that links H$_2$ exposure to H accumulation in GBs, providing information on the early stages of hydrogen-induced degradation.
\end{abstract}


\begin{figure}[h]
\centering
\includegraphics[width=\textwidth]{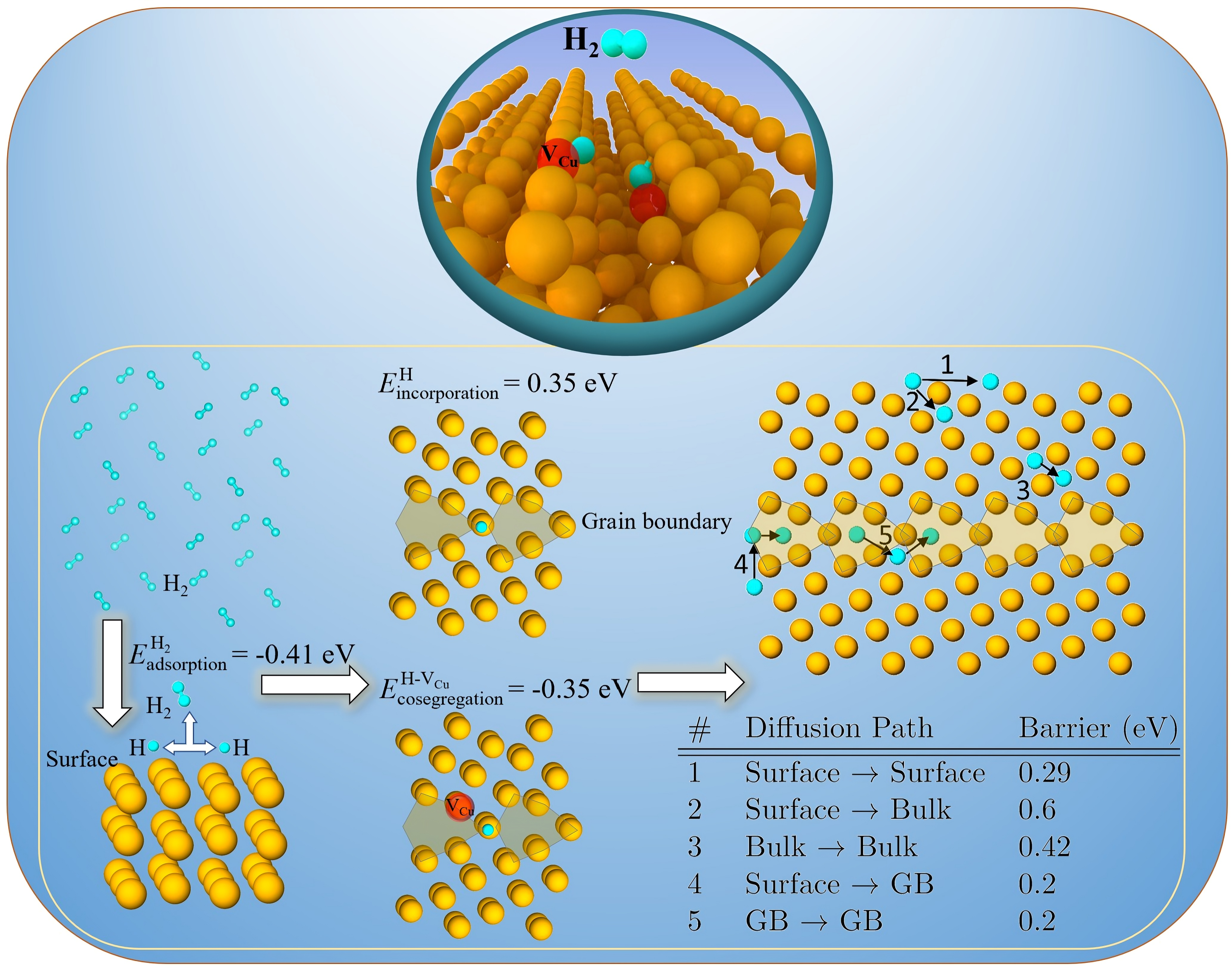}
\end{figure}








\newpage

\section{Introduction \& Background}

Hydrogen (H) affects the mechanical and structural properties of polycrystalline metals, a phenomenon known as hydrogen embrittlement (HE) \cite{robertson2015hydrogen,li2020review}. Despite extensive research, the underlying atomistic mechanisms of HE remain not fully understood, particularly how H interacts with microstructural features, such as grain boundaries (GBs)  \cite{bruck2018hydrogen, gong2020hydrogen}. Of particular concern is the formation of voids and cracks in GBs that are thought to be initiated by the segregation of the H atoms in GBs \cite{fotopoulos2023molecular,yamaguchi2019first}. It is hypothesized that H incorporation accelerates damage processes and structural degradation \cite{polfus2020hydrogen, fotopoulos2024simulation, dadfarnia2019model}.

Copper is widely used in electrical, thermal and structural applications due to its excellent conductivity and mechanical reliability \cite{bhalothia2019h2, bhalothia2019effects, merchant2000metallurgy, fotopoulos2024modeling}. However, voids in Cu-based interconnects and structural components have been reported under mechanical or thermal-induced stresses \cite{kleinbichler2021quantitative}. Experimental observations show that the formation of voids in polycrystalline Cu occurs preferentially in GBs \cite{kleinbichler2021quantitative,moser2022fatigue,arnaud2003evidence}, which are expected to act as vacancy and solute sinks \cite{huang2018uncovering}. In particular, voiding has been shown to occur at moderate temperatures and depends on the intensity and frequency of cyclic stress, even without prolonged thermal exposure \cite{kleinbichler2021quantitative}. Voiding phenomena have been studied in the context of microstructural stress evolution \cite{kleinbichler2021quantitative, moser2019novel, moser2021electropolishing}, however, the specific role of H in this process remains underexplored. 

Although Cu is not a hydride-forming metal, it is known that H can significantly reduce its ductility and fracture resistance \cite{nakahara1988microscopic}. HE mechanisms differ across different metallic systems. In steels, the release of hydrogen during the austenite-martensite transformation can induce local embrittlement, while in Ti and Zr alloys, the trapping of the second-phase particles drives the precipitation of hydride and delays cracking \cite{chen2025hydrogen}. Despite these advances, the atomistic role of hydrogen in face-centered cubic (FCC) Cu remains poorly understood, particularly because Cu exhibits very low bulk H solubility \cite{fotopoulos2023thermodynamic}. Therefore, it is critical to understand the processes initiated by H$_2$ adsorption at Cu surfaces. Previous first-principles calculations have shown that the adsorption energies of H$_2$ on Cu (100) surface range from $-$0.42 to $-$0.91 eV (when using dispersion corrections) and that H$_2$ molecules dissociate with a relatively small barrier of 0.12 eV \cite{alvarez2016hydrogen}. Regions on Cu surfaces with reduced atomic coordination, such as steps or vacancies, are known to enhance molecular and atomic adsorption \cite{mistry2024hydroxyl, jedidi2015generation}. Moreover, grooves are known to form at the intersections of GBs and free surfaces in a polycrystalline material at elevated temperatures \cite{amram2014grain, bentria2017role}, creating undercoordinated sites similar to stepped surfaces. These grooved regions could potentially provide ideal trapping sites for H$_2$ and facilitate subsequent dissociation into atomic H.

Once H atoms are adsorbed on the surface, especially near GBs, they can be directly incorporated into GB interstitial sites. GBs are expected to act as strong traps for H and sinks for Cu vacancies (V$_{Cu}$), setting the stage for the formation of H--vacancy (H--V$_{Cu}$) complexes \cite{momida2013hydrogen, du2020hydrogen, sobola2024exploring}. Previous studies have suggested that hydrogen accumulation in GBs can contribute to decohesion through dislocation-enhanced voiding \cite{robertson2015hydrogen, sobola2024exploring}. Hydrogen-enhanced localized plasticity (HELP) and hydrogen-induced decohesion models describe how solute hydrogen facilitates dislocation activity and interfacial decohesion, ultimately leading to crack nucleation and propagation \cite{robertson2015hydrogen,birnbaum1994hydrogen}. These mechanisms imply that H not only reduces the fracture resistance but also promotes local plasticity and facilitates the local accumulation of vacancies that can accelerate the coalescence of the void \cite{ganchenkova2014effects}. Furthermore, GBs can serve not only as trapping regions, but also as fast diffusion pathways for H transport \cite{lousada2023pathways}, enabling accumulation in critical regions and increasing the likelihood of hydrogen-assisted failure. Thus, the question arises of what is the atomistic mechanism that connects hydrogen exposure to its accumulation at GBs. 

To understand whether hydrogen can promote voiding in polycrystalline Cu, it is essential to link all the aforementioned mechanisms including surface adsorption, segregation, and diffusion of H at the atomic scale. Previous first-principles studies have provided insight into hydrogen energetics and diffusion in specific Cu grain boundary symmetries, particularly $\Sigma$5 and $\Sigma$9, \cite{lousada2023pathways,lousada2022single,huang2023hydrogen}. However, these studies were constrained by small supercells and periodic boundary conditions, which can introduce artificial image interactions and limit the spatial extent of atomic relaxations. Moreover, previous work has treated surfaces, grain boundaries, and bulk regions as isolated systems, without addressing the continuous diffusion pathways that connect them. In this work, we bridge these limitations by combining density functional theory (DFT) and bond-order potential (BOP) simulations within a unified framework that captures surface, grain boundary, and bulk regions simultaneously. The larger BOP supercells (over 900 atoms) eliminate periodic-image artifacts and enable the study of long-range structural relaxations and hydrogen diffusion pathways inaccessible to DFT alone. This hybrid approach allows us to describe, within one consistent model, the sequence of processes from H$_2$ adsorption and dissociation on Cu surfaces to hydrogen incorporation, diffusion, and H–vacancy cosegregation at grain boundaries. 

We start by examining the incorporation and adsorption of H from the gas phase in various crystallographic regions of Cu (bulk, surface, and grain boundaries). Our simulations show that atomic H preferentially adsorbs at undercoordinated Cu surface sites where there are Cu vacancies, in agreement with the literature that stepped and less coordinated surfaces attract adsorbents \cite{mistry2024hydroxyl} and, more specifically, atomic H more strongly \cite{jedidi2015generation}. Since such undercoordinated sites can form in regions where GBs emerge on surfaces \cite{amram2014grain, bentria2017role}, we show that atomic H can be adsorbed on the Cu surface with an adsorption energy of $-$0.24 eV (referenced to the gas phase), which decreases to $-$0.3 eV when a Cu vacancy is present on the surface. Atomic H can be incorporated into GBs with a lower incorporation energy (0.35 eV; referenced to the gas phase) than in bulk Cu (0.68 eV) and further stabilizes Cu vacancies at the GB, lowering the vacancy segregation energy from $-$0.72 eV to $-$0.83 eV (referenced to the bulk). We show that H and Cu vacancies cosegregate in GBs with an energy gain of up to $-$0.8 eV, forming stable H–V$_{Cu}$ complexes. Finally, we investigate the diffusivity of H in various crystallographic regions. H can diffuse from the Cu surface to an emerging GB and along the GB with barriers as low as 0.2 eV, significantly lower than the 0.42 eV barrier in the bulk. These results establish the following atomistic mechanism: atomic H adsorbs strongly on Cu surfaces in the regions where GBs emerge and then diffuses into the GB and accumulates,  stabilizing vacancies and forming H--V$_\textrm{Cu}$ complexes that can potentially seed the formation of voids.

\section{Calculation Methods}

The combination of DFT and BOP methods allows us to use their complementary strengths. Our previous studies have shown that BOPs accurately capture structural relaxations and hydrogen segregation energies in Cu GBs \cite{fotopoulos2024simulation}. However, DFT is required to model electronic structure changes induced by H-vacancy interactions, which BOPs cannot capture. In particular, DFT calculations enable us to investigate charge density transfer and redistribution effects, which are expected to affect the energetics of H--vacancy complexes. Similarly, changes in local charge density due to the presence of H in undercoordinated regions, such as GBs, can be captured by DFT \cite{huang2018uncovering} and are also expected to affect the H--vacancy energy landscape. When it comes to diffusion barrier calculations, the results of our BOP calculations for H diffusion barriers in bulk Cu (0.42 eV) are in excellent agreement with our DFT results (0.4 eV), validating the accuracy of BOPs for large-scale diffusion simulations, particularly at grain boundaries where larger simulation cells are necessary.

\subsection{\label{sec:level3}Computational Details: First-Principles}

DFT calculations of GB are carried out in 76-atom periodic cells (dimensions of 7.27\,{\AA}$\times$8.13\,{\AA}$\times$\newline24.39\,{\AA}) whereas 108-atom cells are used for bulk and surface calculations (10.73\,{\AA}$\times$10.73\,{\AA}$\times$\newline10.73\,{\AA}; 10.73\,{\AA}$\times$10.73\,{\AA}$\times$30.73\,{\AA}). We consider the twin boundary $\Sigma5$(210)[100], which is one of the most widely used GB symmetries in Cu~\cite{huang2018uncovering,huang2019combined,bodlos2023modification,fotopoulos2024first}. In addition, H atoms have been shown to segregate more systematically in $\Sigma5$ GBs in Cu compared to other symmetries (i.e. $\Sigma3$ and $\Sigma11$) \cite{lousada2020hydrogen}. The GB simulation cell is periodically translated into the $x$-, $y$-, and $z$-directions. Along the $z$-direction, a 10\,{\AA} vacuum is added to avoid interactions between periodically translated images. For the 76-atom GB and the 108-atom bulk cells, in line with previous works~\cite{bodlos2023modification,fotopoulos2023thermodynamic,fotopoulos2024simulation,fotopoulos2024first}, converged 5$\times$4$\times$1 and 4$\times$4$\times$4 k-point grids are used, respectively, with an energy cut-off of 450\,eV. In the case of GB cells, because of the added vacuum,  one k-point is used along the $z$-direction. The Cu pseudopotential with 11 valence electrons ($3d^{10} 4s^1$) is used in all calculations. The Vienna Ab Initio Simulation Package (VASP)~\cite{kresse1993ab, kresse1996efficient,kresse1996efficiency} and the Perdew–Burke–Ernzerhof (PBE) generalized gradient approximation functional (GGA)~\cite{perdew1996generalized} are used. A mixture of conjugate gradient (CG)~\cite{er1975iterativecalculationof} and RMM-DIIS~\cite{pulay1980convergence,wood1985new} algorithms is used to minimize the energy with energy and force tolerances of 10$^{-5}$\,eV and 10$^{-2}$\,eV/\AA, respectively.

\subsection{Computational Details: Bond-Order Potentials}

BOP simulations are performed using the Large-Scale Atomic Molecular Massively Parallel Simulator (LAMMPS) code \cite{thompson2022lammps}. The BOP developed by Zhou \textit{et al.} \cite{zhou2015analytical} allows us to model, in addition to pure Cu, the interactions between Cu and H as well as between H atoms. This potential has been reported to accurately reproduce properties for pure FCC Cu, such as stacking fault energy, melting temperature, elastic constants, and surface energies \cite{zhou2015analytical, zhou2016analytical}. Previous work has shown that BOPs provide an accurate representation of the energetic properties and relaxation effects of H interstitials in Cu, demonstrating good agreement with DFT results \cite{fotopoulos2024simulation}.

For computing H diffusion barriers, we employed the nudged elastic band (NEB) method implemented in LAMMPS. The conjugate gradient method was used as the minimizer, with a tolerance for forces set to $10^{-10}$\,eV/{\,\AA}, with a spring constant of 1\,eV/{\,\AA}$^2$. The time step of 10\,fs was chosen in accordance with previous studies, using a timestep approximately ten times larger than the typical 1 fs timestep used for molecular dynamics (MD) simulations. The implementation of NEB in LAMMPS follows refs.\cite{henkelman2000improved,henkelman2000climbing,maras2016global,nakano2008space}.

To ensure a more systematic approach, MD simulations were first performed to identify preferential diffusion pathways by monitoring H migration at elevated temperatures. The identified pathways were then further refined using NEB calculations to obtain accurate energy barriers along the observed trajectories. This approach allows us to first explore the dynamic behavior of hydrogen diffusion, and then quantify the associated barriers for the most probable migration paths. We performed the MD simulations in a canonical ensemble (NPT) to control the temperature and pressure. The timestep for all MD simulations was set to 1 fs, and a Nosé--Hoover thermostat and barostat were used with damping parameters set at 1 ps to ensure proper equilibration. The velocities were assigned using a Maxwell--Boltzmann distribution corresponding to a target temperature of 300 K. During the MD runs, the temperature was gradually increased from 0 K to the target temperature (700 K) over 100 ps.

\subsection{Calculation of Energetic Parameters}

Segregation and migration energies are two fundamental parameters for classifying GBs as trapping locations for solute diffusion \cite{ding2024hydrogen}.
 Segregation energies are calculated as:

\begin{equation}
    E_{seg}=E_{GB+X}-E_{Bulk+X},
    \label{eq:5.1}
    \end{equation} 
    
\noindent{where} $E_{GB+X}$ and $E_{Bulk+X}$ are the energies with segregants in the grain boundary and bulk regions, respectively. Negative segregation energies indicate a preference for solutes to segregate at the grain boundary. To determine the segregation energies related to interstitial impurities, we consider only octahedral interstitial sites, since previous theoretical studies have established that H atoms prefer to occupy octahedral over tetrahedral sites in FCC Cu \cite{fotopoulos2024simulation}.

Cosegregation at the GB is a phenomenon in which specific elements, including vacancies, preferentially locate at the interfaces between grains in polycrystalline materials. This process is driven by the reduction in the system's free energy, leading to a distinct arrangement of segregants (H atoms) and vacancies at the GB. It can be characterized by the cosegregation energy, which determines the stability and distribution of these species at the GB. The cosegregation energy ($E_{{coseg}}$) is calculated as:
\begin{equation}
E_{\text{coseg}} = E_{\text{seg}}^{\text{H+V}_{\text{Cu}}} - \left(E_{\text{seg}}^{\text{H}} + E_{\text{seg}}^{\text{V}_{\text{Cu}}}\right),
\end{equation}
where $E_{\text{seg}}^{\text{H+V}_{\text{Cu}}}$ is the segregation energy of H and a Cu vacancy together at the GB, while $E_{\text{seg}}^{\text{H}}$ and $E_{\text{seg}}^{\text{V}_{\text{Cu}}}$ are the segregation energies of H and the Cu vacancy individually (each referenced to the bulk). A negative $E_{\text{coseg}}$ indicates that the simultaneous segregation of H and the vacancy at the GB is more favorable than the sum of their independent segregation energies, i.e., that they preferentially form a complex at the GB. Finally, the H incorporation/adsorption energies are computed using the formula: 

\begin{equation}
E_{\text{inc/ads}} = E_{H} - (E_{pr}) - \mu_{H},
\end{equation}
where $E_{H}$ and $E_{pr}$ are the energies of simulation cells with interstitial H (bulk, surface, or GB cell, in the case of the surface, we compute the adsorption energy) and pristine (H-free), respectively, with $\mu_{H}$ half of the DFT-computed energy of an isolated H$_2$ molecule. For adsorption energies, D2-corrected PBE is used \cite{grimme2006semiempirical}.

\section{Results}

\subsection{DFT Results: Hydrogen Adsorption and Vacancy Segregation in Bulk, Surface, and GB}

\begin{figure}[h]
\centering
\includegraphics[width=\textwidth]{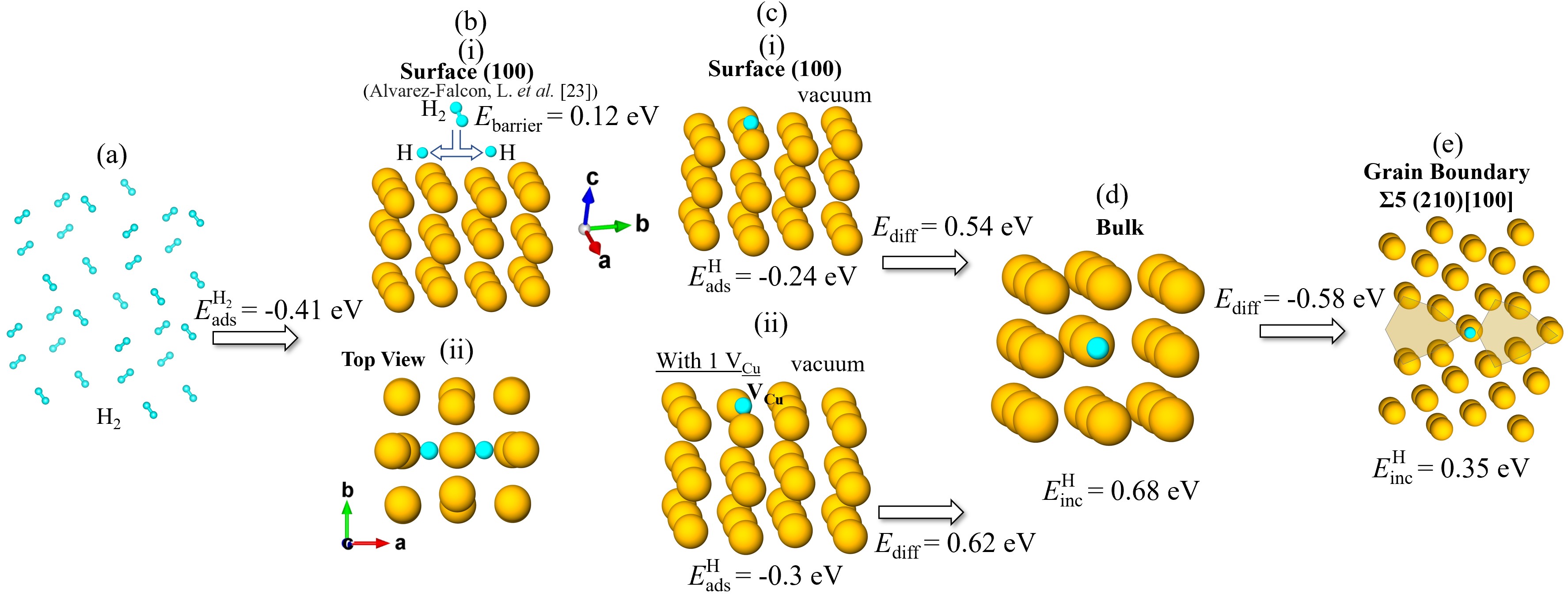}
\caption{Hydrogen adsorption, dissociation, and relative energetics across different crystallographic environments in Cu. 
(a) Molecular hydrogen (H$_2$) in the gas phase. 
(b) Dissociation of adsorbed H$_2$ on Cu (100): the reported barrier for this process is 0.12 eV~\cite{alvarez2016hydrogen}. Panel (ii) shows the top view after dissociation.
(c) Atomic H adsorption on Cu (100): (i) pristine surface with $E^{\mathrm{H}}_{\mathrm{ads}}=-0.24$ eV; (ii) surface containing a Cu vacancy with stronger adsorption, $E^{\mathrm{H}}_{\mathrm{ads}}=-0.30$ eV. The arrows labelled $E_{\mathrm{diff}}$ indicate energy differences between the two states: from H on the pristine surface to H in bulk, $E_{\mathrm{diff}}=+0.54$ eV; from H in a surface vacancy to H in bulk, $E_{\mathrm{diff}}=+0.62$ eV.
(d) Incorporation of H in bulk Cu $E^{\mathrm{H}}_{\mathrm{inc}}=0.68$ eV.
(e) Incorporation of H at the $\Sigma 5$(210)[100] grain boundary (GB) is less unfavorable, $E^{\mathrm{H}}_{\mathrm{inc}}=0.35$ eV. The arrow from bulk to GB shows the energetic offset $E_{\mathrm{diff}}=-0.58$ eV, i.e., H in GB configuration is lower in energy than H in bulk by 0.58 eV.
All adsorption/incorporation energies are referenced to molecular H$_2$ in the gas phase; $E_{\mathrm{diff}}$ values quantify final–initial total-energy differences between the indicated configurations.}
\label{fig:fig1}
\end{figure}

We start by examining the energetics of hydrogen adsorption from the gas phase (H$_2$) and its incorporation into Cu surfaces, bulk, and grain boundaries (Figure~\ref{fig:fig1}). The adsorption energy of H$_2$ on the Cu (100) surface is $E$\textsubscript{ads} = –0.41 eV (Figure~\ref{fig:fig1}(a)). Once adsorbed (Figure~\ref{fig:fig1}(b)), H$_2$ has been reported to dissociate into atomic H on Cu (100) surface with a dissociation barrier of 0.12 eV \cite{alvarez2016hydrogen}. Our DFT calculations show that the atomic H adsorption is moderately favorable on the pristine Cu (100) surface (Figure~\ref{fig:fig1}(c)(i)), with an adsorption energy of $E$\textsubscript{ads} = $-$0.24 eV (reference to the gas phase). This adsorption becomes slightly more favorable in the presence of a Cu surface vacancy, where H is incorporated inside the vacancy, with $E$\textsubscript{ads} = $-$0.3 eV (Figure~\ref{fig:fig1}(c)(ii)). The latter indicates that intrinsic surface defects improve the strength of H binding, consistent with previous studies on undercoordinated surfaces and dopant adsorption \cite{mistry2024hydroxyl, jedidi2015generation}. In the bulk (Figure~\ref{fig:fig1}(d)), H incorporation from the gas phase is thermodynamically unfavorable (E\textsubscript{inc} = 0.68 eV), in agreement with the known low solubility of H in Cu \cite{fotopoulos2023thermodynamic}. However, the incorporation energy is significantly reduced at the $\Sigma$5(210)[100] grain boundary (Figure~\ref{fig:fig1}(e)), where $E$\textsubscript{inc} = 0.35 eV, supporting the role of GBs as preferential trapping sites for H.

\begin{figure}[h]
\centering
\includegraphics[width=\textwidth]{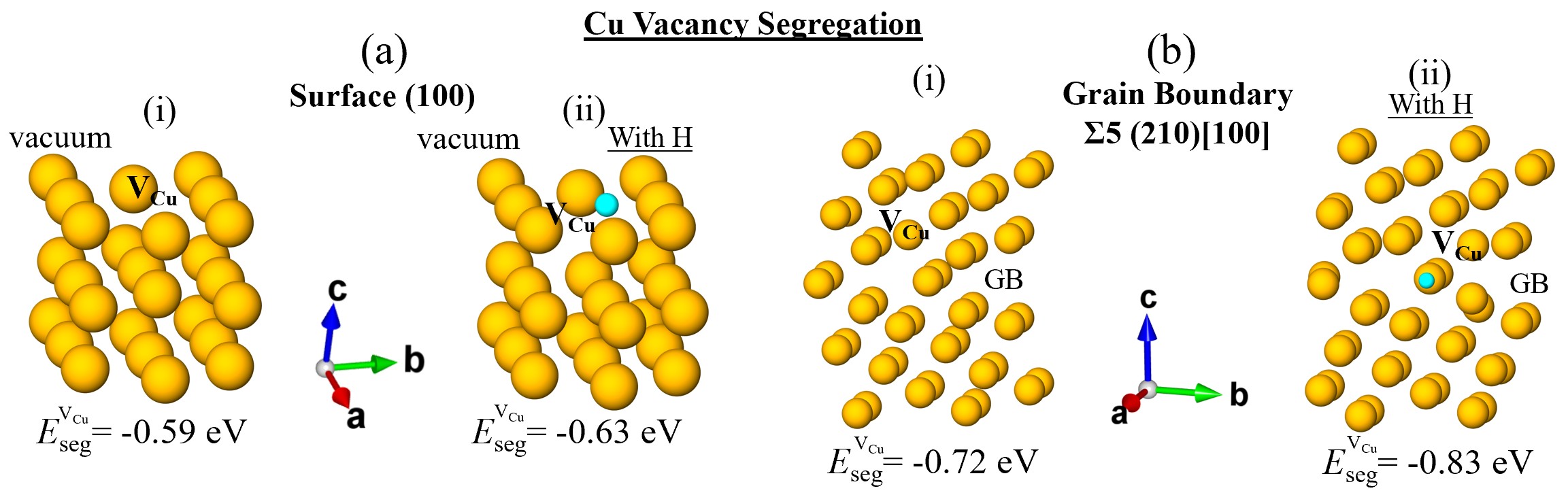}
\caption{Copper vacancy segregation energetics at the Cu (100) surface and the $\Sigma 5$(210)[100] GB, with and without the presence of hydrogen. Segregation energies ($E_{\text{seg}}^{\text{V}_{\text{Cu}}}$) are referenced to a vacancy located in bulk Cu.
(a) (i) Surface vacancy segregation at the pristine Cu (100) surface is energetically favorable with $E_{\text{seg}}^{\text{V}_{\text{Cu}}} = -0.59$ eV. (ii) The presence of hydrogen slightly enhances vacancy segregation to $-0.63$ eV. (b) (i) Vacancy segregation at the $\Sigma 5$(210)[100] GB is more favorable ($E_{\text{seg}}^{\text{V}_{\text{Cu}}} = -0.72$ eV) compared to the surface. (ii) The presence of hydrogen at the GB further lowers the energy to $-0.83$ eV. The negative segregation energies indicate a clear thermodynamic driving force for vacancy accumulation at grain boundaries and surfaces, enhanced by hydrogen incorporation.}
\label{fig:fig2}
\end{figure}

Next, we investigate the segregation of Cu vacancies at the surface and grain boundary, as summarized in Figure~\ref{fig:fig2}. Segregation energies are calculated relative to a vacancy located in the bulk. At the Cu (100) surface (Figure~\ref{fig:fig2}(a)), vacancy segregation is favorable ($E$\textsubscript{seg} = $-$0.59 eV) and becomes slightly more favorable (–0.63 eV) in the presence of hydrogen. At the $\Sigma$5(210)[100] grain boundary (Figure~\ref{fig:fig2}(b)), segregation is even more favorable ($E$\textsubscript{seg} = $-$0.72 eV), and further enhanced by hydrogen (–0.83 eV). Thus, vacancies preferentially segregate to GBs over surfaces and bulk and the presence of hydrogen further stabilizes these vacancies at GBs. Thus far, these results point to a synergistic interaction between hydrogen and Cu vacancies at GBs, facilitating the formation of stable H–V\textsubscript{Cu} complexes. This synergy may provide a thermodynamic driving force for early-stage void nucleation at grain boundaries.

\subsection{DFT Results: Interaction Between H Interstitials and Cu Vacancies at the GB}

\begin{figure}[t]\vspace*{4pt}
\centering
\includegraphics[width=\textwidth]{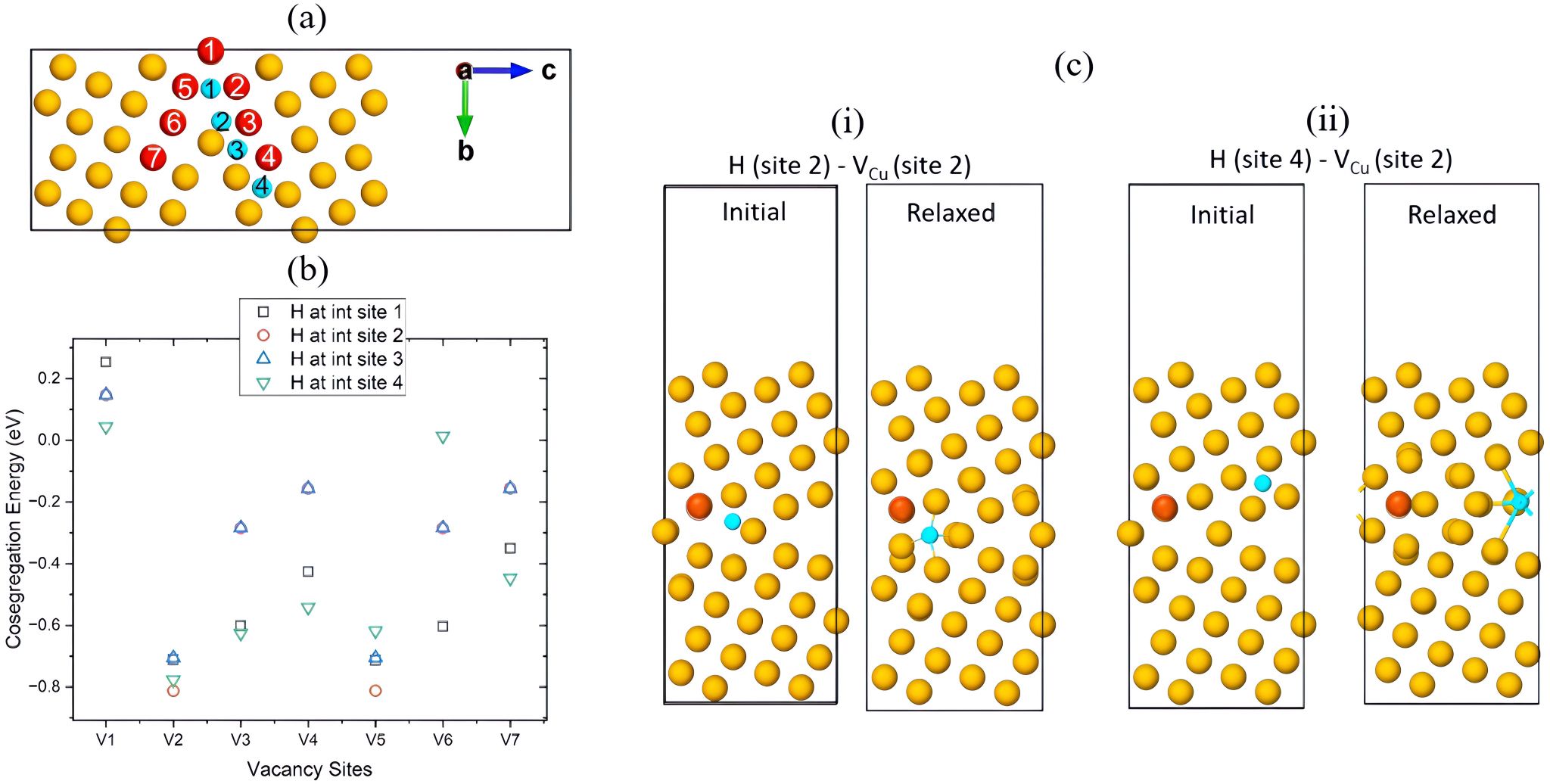}\\
\caption{(a) The 76-atom Cu $\Sigma5$(210)[100] GB simulation cell showing the examined segregation sites for Cu vacancies (red) and hydrogen interstitial atoms (cyan). (b) Computed cosegregation energies for H interstitials occupying different sites (1--4) interacting with Cu vacancies located at various GB sites (V1--V7). Negative energies indicate favorable cosegregation at the GB. (c) Atomic configurations of the two representative H--V$_\text{Cu}$ complexes: (i) H initially at interstitial site 2 cosegregating with a Cu vacancy at site 2; (ii) H initially at interstitial site 4 cosegregating with a Cu vacancy at site 2. Initial and relaxed configurations illustrate significant local atomic rearrangements upon relaxation.}
\label{fig:Voidsdissociation}
\end{figure}

Having established the preferential incorporation sites for hydrogen and segregation energies of vacancies at surfaces and GBs, we now turn to their mutual co-existence at GBs. Previous theoretical studies \cite{huang2018uncovering,fotopoulos2024simulation}, in agreement with the results presented in the previous section, have highlighted that interstitial H atoms prefer to segregate in GB compared to the bulk of Cu. Therefore, we are interested in the formation of H--V$_\textrm{Cu}$ complexes at the GB. To understand this, we calculate the cosegregation energies of H and a Cu vacancy in the symmetric $\Sigma5$ GB (shown in Figure \ref{fig:Voidsdissociation}(a)). The computed energies are summarized in Figure \ref{fig:Voidsdissociation}(b). The plot shows that, in the case where the interstitial H and a vacancy are introduced simultaneously at the GB while retaining a close distance, their interaction leads to a significantly higher gain in energy (up to $-$0.8 eV) than the presence of the two entities at the GB individually. This substantial gain in energy arises from two synergistic effects: (i) interstitial H incorporation at the GB induces local lattice distortion and lowers nearby atomic coordination, effectively lowering the energetic cost of forming a Cu vacancy; and (ii) the Cu vacancy itself creates an undercoordinated region with additional free volume, which is favorable for H incorporation, consistent with our previous findings that H preferentially incorporates in such open, low-coordination environments. Thus, H--V$_\textrm{Cu}$ complexes are significantly more energetically favored to form at the GB of the crystal. The initial and fully relaxed configurations for the lowest energy H--V$_\textrm{Cu}$ complex, highlighting this preferential interaction, are depicted in Figure \ref{fig:Voidsdissociation}(c)(i). Even when the vacancy and H are initially far apart (Figure \ref{fig:Voidsdissociation}(c)(ii)), relaxation leads to local distortion that drives the two defects closer together, still resulting in a significant energy gain ($-$0.79 eV).

\subsection{BOP Results: H Interstitial Diffusion at Elevated Temperatures}

\begin{figure}[t]\vspace*{4pt}
\centering
\includegraphics[width=0.9\textwidth]{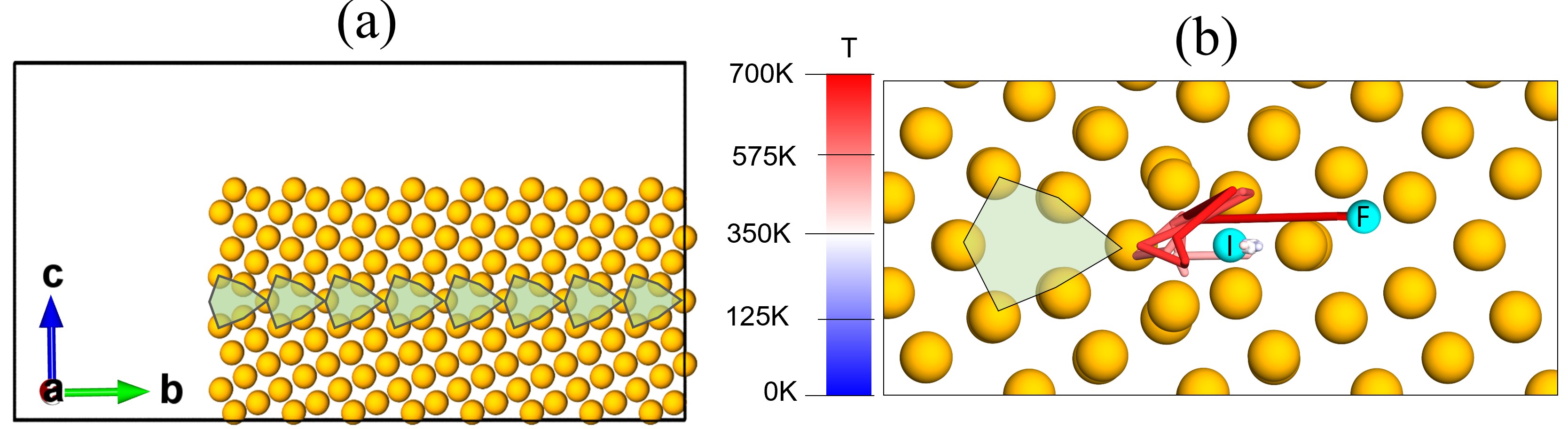}
\caption{\label{fig:MDtrajectory} 
(a) Simulation cell used for MD and NEB barrier calculations, comprising 932 atoms with two free surfaces and a central $\Sigma5$(210)[100] GB.
(b) MD trajectories illustrating hydrogen diffusion within Cu at 700 K. Trajectories are colored according to temperature, highlighting preferential H migration paths along the grain boundary. Labels `I' (initial) and `F' (final) indicate start and end positions of a hydrogen atom during the MD simulation.}
\end{figure}

Although the tendency for Cu vacancies to form and segregate at GBs and interfaces is well documented \cite{bodlos2023modification,fotopoulos2023structure}, the specific mechanisms that drive the accumulation and transport of H in these regions are less understood. To address this, we employed simulations that extend beyond the computational limitations of DFT by using BOPs. To identify energetically favorable diffusion pathways before calculating diffusion barriers, we performed MD annealing simulations at temperatures up to 700 K. Figure \ref{fig:MDtrajectory}(a) illustrates the GB model used for these simulations, which consists of 932 atoms and features a symmetric $\Sigma5$(210)[100] GB, along with two free surfaces, the (210) and (100) crystallographic planes. This model enables us to study the diffusion of H in the bulk, surface, and GB regions but also to capture surface-to-GB-to-bulk diffusion pathways and long-range structural relaxations that are inaccessible to smaller DFT cells. The diffusion trajectories extracted from the MD simulations, shown in Figure \ref{fig:MDtrajectory}(b), demonstrate that at temperatures above 500\,K, H starts to migrate within the GB, eventually moving from the center of symmetry of a $\Sigma5$ kite to an octahedral site in a neighboring kite at temperatures close to 700\,K. Subsequently, H can be absorbed into the center of symmetry of adjacent $\Sigma5$ kites without any discernible energy barrier, indicating efficient transport pathways within the GB network. Interestingly, the migration pattern observed during the MD simulations, where H hops between adjacent $\Sigma5$ kite centers via intermediate octahedral sites, is very similar to the diffusion pathway reported in previous DFT studies \cite{lousada2023pathways}.

\subsection{BOP Results: H Interstitial Diffusion Barriers}

\begin{figure}[t]\vspace*{4pt}
\centering
\includegraphics[width=0.9\textwidth]{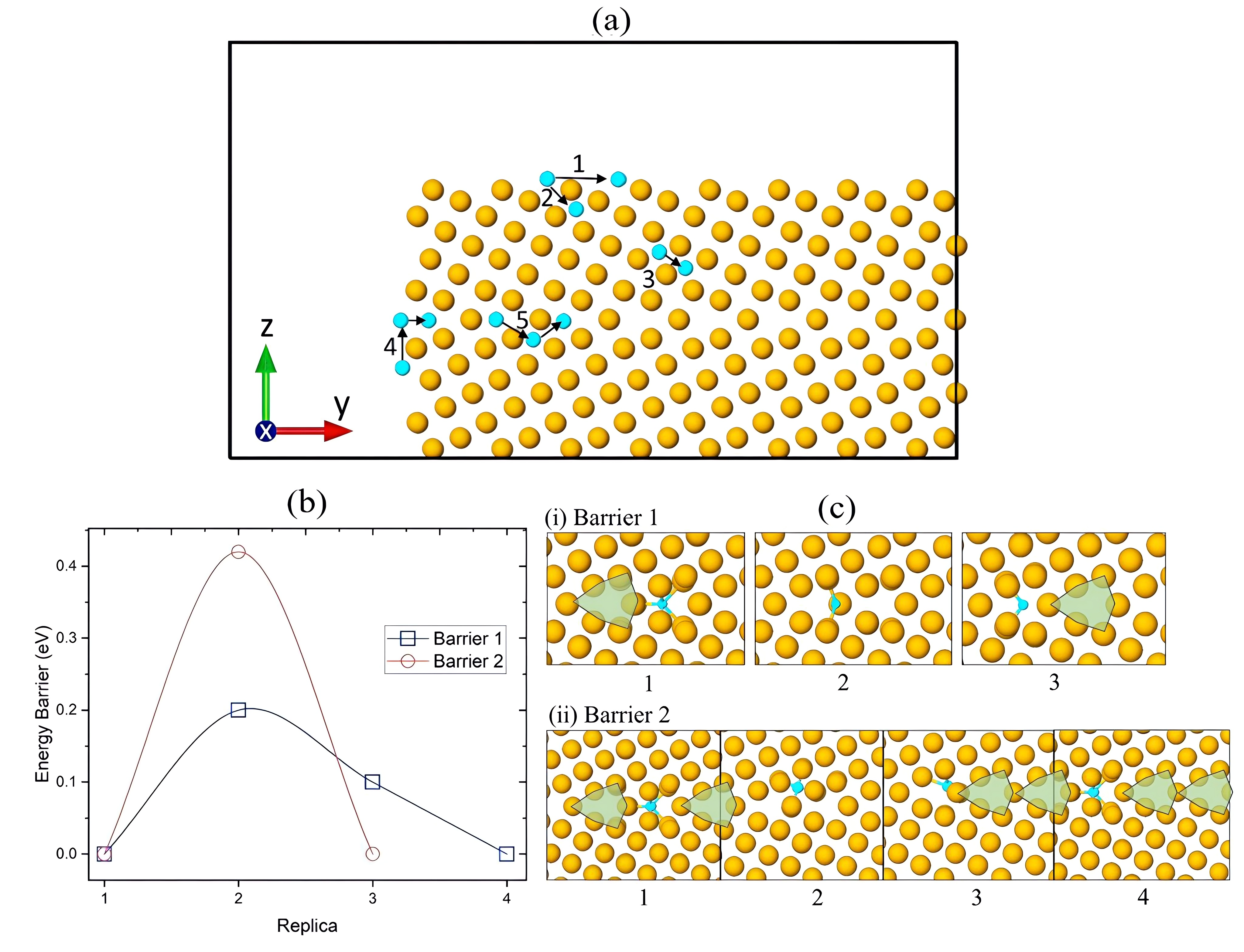}\\
\caption{\label{fig:epsart} (a) The various diffusion paths examined for H. (b) Diffusion barriers calculated through NEB and BOP for H in two potential diffusion paths within the GBs. (c) (i),(ii) Configurations of the images (replicas) during the diffusion process for the two paths, demonstrating the intermediate states that H assumes as it migrates through the GB.}
\label{fig:6}
\end{figure}

Building on the insights obtained from the MD simulations, which identified the most favorable diffusion pathway for H within the GB, we calculated the associated diffusion barriers.  Figure \ref{fig:6}(a) illustrates the various diffusion pathways examined in our study. Hydrogen diffusion along the surface (path 1) occurs with a barrier of 0.29 eV, indicating low barrier H surface diffusion. However, the barrier for migration from the surface into the bulk (path 2) increases to 0.6 eV, reflecting the resistance to H penetration into the more densely packed bulk structure. Within the bulk (path 3), the calculated barrier for diffusion between equivalent octahedral sites is 0.42 eV, consistent with previous findings~\cite{katz1971diffusion,zhou2013dissolution}. When H diffuses from a hollow site on the surface to the center of symmetry of a $\Sigma5$ kite within the GB (path 4), a significantly lower barrier of 0.2 eV is observed, suggesting that the GB serves as a preferential absorption site for H atoms. Once inside the GB (path 5), H can migrate efficiently between adjacent $\Sigma5$ kites with an equally low barrier of 0.2 eV, confirming the role of the GB as a fast diffusion channel. The diffusion barrier we compute for hydrogen migration within the $\Sigma5$ GB (0.20 eV) is in excellent agreement with previous DFT calculations \cite{lousada2023pathways}. This confirms that smaller DFT cells are sufficient to capture the local energetics of H diffusion along individual GB cores. However, such limited cells cannot resolve the coupling between distinct crystallographic regions. Unlike earlier studies restricted to intra-GB motion of interstitial H, our model explicitly resolves surface-to-GB and GB-to-bulk transitions within the same simulation cell, revealing a continuous diffusion pathway.

Figure \ref{fig:6}(b) shows the diffusion barriers for the two H diffusion paths examined along the GB. In the first path (Figure \ref{fig:6}(c)(i)), H diffuses along a straight trajectory that runs through the central axis of the GB, following the line connecting the centers of symmetry of adjacent $\Sigma5$ kites. However, when H diffuses through the coincident site of the coincidence site lattice (CSL), due to the local coherence of the GB, the barrier is close to that of H in the bulk (namely 0.4\,eV). Similarly to the diffusion path shown in Figure \ref{fig:MDtrajectory}(b), H diffuses from the center of symmetry of a $\Sigma5$ kite to an adjacent octahedral site with a barrier of 0.2 eV (Figure \ref{fig:6}(c)(ii)). Subsequently, H can be absorbed into the center of symmetry of a nearby $\Sigma$5 kite with zero barrier. This process demonstrates a low barrier mechanism for the diffusion of H along the GB and the formation of H–V\textsubscript{Cu} complexes, considering the significantly lower formation energies of Cu vacancies in the GB \cite{fotopoulos2023structure}. Table 1 summarizes the diffusion barriers computed for H for the various pathways. The significantly lower barriers in GB suggest that they can serve as effective pathways for H transport, which could play a critical role in the accumulation of H at these interfaces, potentially leading to embrittlement.

\begin{table}[t]
\centering
\small
\caption{Lowest diffusion barriers for each path computed with NEB and BOP.}
\label{tab:barriers}
\begin{tabular}{c l c}
\toprule
\# & Diffusion path & Barrier (eV) \\
\midrule
1 & Surface $\to$ Surface & 0.29 \\
2 & Surface $\to$ Bulk    & 0.60 \\
3 & Bulk $\to$ Bulk       & 0.42 \\
4 & Surface $\to$ GB      & 0.20 \\
5 & GB $\to$ GB           & 0.20 \\
\bottomrule
\end{tabular}
\end{table}

\section{Discussion and Conclusions}

On the basis of the results presented in this paper, we can formulate an atomistic mechanism of initial processes leading to hydrogen-induced degradation in Cu. In the first step, H$_2$ is adsorbed on the Cu surface ($E$\textsubscript{ads} = $-$0.41 eV) followed by dissociation to atomic H with a reported low barrier of 0.12 eV \cite{alvarez2016hydrogen}. On pristine Cu surfaces, atomic H binds with an adsorption energy of $-$0.24 eV, while the presence of a surface vacancy slightly enhances this binding to $-$0.30 eV. Undercoordinated surface sites favor atomic H adsorption and retention. Importantly, GB grooving \cite{amram2014grain} could enable a direct incorporation pathway for atomic hydrogen from the gas phase into the GB core, bypassing the need for bulk diffusion.

Once atomic hydrogen enters the GB, it encounters a thermodynamically favorable environment. The hydrogen incorporation energy at the GB (0.35~eV) from the gas phase is substantially lower than in bulk Cu (0.68~eV), supporting the GB’s role as a hydrogen sink. Cu vacancies also segregate favorably to GBs ($E_{\text{seg}}^{\text{V}_{\text{Cu}}} = -0.72$~eV; favorable segregation at the GB compared to bulk), and the presence of H further stabilizes this segregation ($-0.83$~eV). Our results thus indicate strong driving forces for the formation of H--V$_\text{Cu}$ complexes in the GB, with cosegregation energy gains of up to $-0.8$~eV. Crucially, hydrogen diffuses within the GB through low energy barriers ($\sim$0.2~eV), enabling fast redistribution across extended GB networks. This is in contrast to bulk Cu, where diffusion barriers are significantly higher (0.42~eV). Fast GB diffusion allows hydrogen to accumulate at stress concentration regions, where it can further exacerbate damage through HELP-mediated dislocation mobility. 

Our results for FCC Cu suggest a distinct pathway for hydrogen embrittlement compared to steels and hydride-forming alloys. Whereas in steels embrittlement is often linked to hydrogen release during phase transformations, and in Ti/Zr alloys to hydride precipitation \cite{chen2025hydrogen}, in Cu the critical process involves the fast diffusion of hydrogen along grain boundaries and the stabilization of vacancies through the formation of H–V\textsubscript{Cu} complexes. These complexes provide a thermodynamic driving force for void nucleation, while low diffusion barriers along GBs enable rapid redistribution of hydrogen to stress-concentrated regions. Thus, while the specific microstructural traps and release mechanisms differ across systems, our findings reinforce the unifying principle that hydrogen-assisted degradation originates from the coupling of hydrogen transport, trapping, and microstructural instability. These distinctions underscore the importance of studying FCC Cu separately as its hydrogen transport and trapping behavior cannot be directly inferred from BCC/FCC  alloys or other FCC metals. In contrast to prior DFT-only work, our combined DFT–BOP framework captures the coupling between surface adsorption, GB trapping, and vacancy cosegregation within a single, continuous simulation cell. This eliminates artificial constraints from small periodic GB models. The inclusion of both surfaces and GBs allows us to reproduce realistic diffusion pathways and energetics linking H$_2$ surface adsorption to H accumulation at the GB.

In conclusion, these findings provide an atomistic understanding of how hydrogen can contribute to void nucleation and embrittlement in polycrystalline Cu: from preferential surface adsorption and incorporation to thermodynamically driven segregation and fast diffusion within GBs. Our results highlight the role of GBs not only as trapping sites but also as efficient pathways for hydrogen redistribution and accumulation, underscoring the need for experimental validation to determine the precise role of hydrogen in voiding in Cu. Voids and vacancy clusters can be detected using microscopy techniques, such as scanning electron microscopy (SEM) and transmission electron microscopy (TEM), which allow for high-resolution imaging of microstructural changes in the material. Furthermore, focused ion beam (FIB) techniques could be employed to cross-sectionalize the samples and analyze the void distributions in the GBs with greater precision \cite{moser2022fatigue}. 

Moving forward, our computed diffusion barriers can serve as critical inputs for kinetic Monte Carlo (KMC) simulations, enabling long-time-scale modeling of hydrogen accumulation and defect evolution under realistic conditions. Previous KMC studies have successfully modeled H–vacancy interactions in metals \cite{williams2023accelerating} and the mobility of hydrogen on metallic surfaces using machine-learned potentials \cite{steffen2024hydrogen}. Furthermore, our atomistic diffusion barriers and trapping energetics provide quantitative input parameters for higher-scale modeling of hydrogen embrittlement. In particular, phase field formulations that couple hydrogen transport, stress, and fracture \cite{navidtehrani2025generalised,kristensen2020applications} can directly incorporate such data through the definition of diffusion coefficients and trap kinetics. Integrating our atomic-scale insights with these continuum models represents a promising future direction to establish a multiscale framework for predicting hydrogen-assisted degradation in Cu.

\section*{Acknowledgements}
\noindent{A.L.S. acknowledges funding by EPSRC (grant EP/P013503/1). V.F. would like to acknowledge funding by EPSRC (grant EP/L015862/1) as part of the CDT in molecular modelling and materials science. Computational resources on ARCHER2 (http://www.archer2.ac.uk) were provided via our membership of the UK's HPC Materials Chemistry Consortium, which is funded by EPSRC (EP/X035859). This work used the UK Materials and Molecular Modelling Hub for computational resources, MMM Hub, which is partially funded by EPSRC (EP/T022213/1, EP/W032260/1 and EP/P020194/1).}


\newpage

\bibliography{cas-refs}

\providecommand{\latin}[1]{#1}
\makeatletter
\providecommand{\doi}
  {\begingroup\let\do\@makeother\dospecials
  \catcode`\{=1 \catcode`\}=2 \doi@aux}
\providecommand{\doi@aux}[1]{\endgroup\texttt{#1}}
\makeatother
\providecommand*\mcitethebibliography{\thebibliography}
\csname @ifundefined\endcsname{endmcitethebibliography}  {\let\endmcitethebibliography\endthebibliography}{}
\begin{mcitethebibliography}{63}
\providecommand*\natexlab[1]{#1}
\providecommand*\mciteSetBstSublistMode[1]{}
\providecommand*\mciteSetBstMaxWidthForm[2]{}
\providecommand*\mciteBstWouldAddEndPuncttrue
  {\def\EndOfBibitem{\unskip.}}
\providecommand*\mciteBstWouldAddEndPunctfalse
  {\let\EndOfBibitem\relax}
\providecommand*\mciteSetBstMidEndSepPunct[3]{}
\providecommand*\mciteSetBstSublistLabelBeginEnd[3]{}
\providecommand*\EndOfBibitem{}
\mciteSetBstSublistMode{f}
\mciteSetBstMaxWidthForm{subitem}{(\alph{mcitesubitemcount})}
\mciteSetBstSublistLabelBeginEnd
  {\mcitemaxwidthsubitemform\space}
  {\relax}
  {\relax}

\bibitem[Robertson \latin{et~al.}(2015)Robertson, Sofronis, Nagao, Martin, Wang, Gross, and Nygren]{robertson2015hydrogen}
Robertson,~I.~M.; Sofronis,~P.; Nagao,~A.; Martin,~M.~L.; Wang,~S.; Gross,~D.; Nygren,~K. {Hydrogen embrittlement understood}. \emph{Metallurgical and Materials Transactions A} \textbf{2015}, \emph{46}, 2323--2341\relax
\mciteBstWouldAddEndPuncttrue
\mciteSetBstMidEndSepPunct{\mcitedefaultmidpunct}
{\mcitedefaultendpunct}{\mcitedefaultseppunct}\relax
\EndOfBibitem
\bibitem[Li \latin{et~al.}(2020)Li, Ma, Zhang, Akiyama, Wang, and Song]{li2020review}
Li,~X.; Ma,~X.; Zhang,~J.; Akiyama,~E.; Wang,~Y.; Song,~X. {Review of hydrogen embrittlement in metals: Hydrogen diffusion, hydrogen characterization, hydrogen embrittlement mechanism and prevention}. \emph{Acta Metallurgica Sinica (English Letters)} \textbf{2020}, \emph{33}, 759--773\relax
\mciteBstWouldAddEndPuncttrue
\mciteSetBstMidEndSepPunct{\mcitedefaultmidpunct}
{\mcitedefaultendpunct}{\mcitedefaultseppunct}\relax
\EndOfBibitem
\bibitem[Br{\"u}ck \latin{et~al.}(2018)Br{\"u}ck, Schippl, Schwarz, Christ, Fritzen, and Weihe]{bruck2018hydrogen}
Br{\"u}ck,~S.; Schippl,~V.; Schwarz,~M.; Christ,~H.-J.; Fritzen,~C.-P.; Weihe,~S. {Hydrogen embrittlement mechanism in fatigue behavior of austenitic and martensitic stainless steels}. \emph{Metals} \textbf{2018}, \emph{8}, 339\relax
\mciteBstWouldAddEndPuncttrue
\mciteSetBstMidEndSepPunct{\mcitedefaultmidpunct}
{\mcitedefaultendpunct}{\mcitedefaultseppunct}\relax
\EndOfBibitem
\bibitem[Gong \latin{et~al.}(2020)Gong, Nutter, Rivera-Diaz-Del-Castillo, and Rainforth]{gong2020hydrogen}
Gong,~P.; Nutter,~J.; Rivera-Diaz-Del-Castillo,~P.; Rainforth,~W. {Hydrogen embrittlement through the formation of low-energy dislocation nanostructures in nanoprecipitation-strengthened steels}. \emph{Science Advances} \textbf{2020}, \emph{6}, eabb6152\relax
\mciteBstWouldAddEndPuncttrue
\mciteSetBstMidEndSepPunct{\mcitedefaultmidpunct}
{\mcitedefaultendpunct}{\mcitedefaultseppunct}\relax
\EndOfBibitem
\bibitem[Fotopoulos \latin{et~al.}(2023)Fotopoulos, O'Hern, and Shluger]{fotopoulos2023molecular}
Fotopoulos,~V.; O'Hern,~C.~S.; Shluger,~A.~L. {Molecular dynamics simulations of the thermal evolution of voids in Cu bulk and grain boundaries}. TMS Annual Meeting \& Exhibition. 2023; pp 1001--1010\relax
\mciteBstWouldAddEndPuncttrue
\mciteSetBstMidEndSepPunct{\mcitedefaultmidpunct}
{\mcitedefaultendpunct}{\mcitedefaultseppunct}\relax
\EndOfBibitem
\bibitem[Yamaguchi \latin{et~al.}(2019)Yamaguchi, Ebihara, Itakura, Tsuru, Matsuda, and Toda]{yamaguchi2019first}
Yamaguchi,~M.; Ebihara,~K.-I.; Itakura,~M.; Tsuru,~T.; Matsuda,~K.; Toda,~H. {First-principles calculation of multiple hydrogen segregation along aluminum grain boundaries}. \emph{Computational Materials Science} \textbf{2019}, \emph{156}, 368--375\relax
\mciteBstWouldAddEndPuncttrue
\mciteSetBstMidEndSepPunct{\mcitedefaultmidpunct}
{\mcitedefaultendpunct}{\mcitedefaultseppunct}\relax
\EndOfBibitem
\bibitem[Polfus \latin{et~al.}(2020)Polfus, L{\o}vvik, Bredesen, and Peters]{polfus2020hydrogen}
Polfus,~J.~M.; L{\o}vvik,~O.~M.; Bredesen,~R.; Peters,~T. {Hydrogen induced vacancy clustering and void formation mechanisms at grain boundaries in palladium}. \emph{Acta Materialia} \textbf{2020}, \emph{195}, 708--719\relax
\mciteBstWouldAddEndPuncttrue
\mciteSetBstMidEndSepPunct{\mcitedefaultmidpunct}
{\mcitedefaultendpunct}{\mcitedefaultseppunct}\relax
\EndOfBibitem
\bibitem[Fotopoulos and Shluger(2024)Fotopoulos, and Shluger]{fotopoulos2024simulation}
Fotopoulos,~V.; Shluger,~A.~L. {Simulation of mechanical effects of hydrogen in bicrystalline Cu using DFT and bond order potentials}. \emph{Procedia Structural Integrity} \textbf{2024}, \emph{52}, 356--3657\relax
\mciteBstWouldAddEndPuncttrue
\mciteSetBstMidEndSepPunct{\mcitedefaultmidpunct}
{\mcitedefaultendpunct}{\mcitedefaultseppunct}\relax
\EndOfBibitem
\bibitem[Dadfarnia \latin{et~al.}(2019)Dadfarnia, Martin, Moore, Orwig, and Sofronis]{dadfarnia2019model}
Dadfarnia,~M.; Martin,~M.~L.; Moore,~D.~E.; Orwig,~S.~E.; Sofronis,~P. A model for high temperature hydrogen attack in carbon steels under constrained void growth. \emph{International Journal of Fracture} \textbf{2019}, \emph{219}, 1--17\relax
\mciteBstWouldAddEndPuncttrue
\mciteSetBstMidEndSepPunct{\mcitedefaultmidpunct}
{\mcitedefaultendpunct}{\mcitedefaultseppunct}\relax
\EndOfBibitem
\bibitem[Bhalothia \latin{et~al.}(2019)Bhalothia, Lin, Yan, Yang, and Chen]{bhalothia2019h2}
Bhalothia,~D.; Lin,~C.-Y.; Yan,~C.; Yang,~Y.-T.; Chen,~T.-Y. {H2 reduction annealing induced phase transition and improvements on redox durability of Pt cluster-decorated Cu@ Pd electrocatalysts in oxygen reduction reaction}. \emph{ACS Omega} \textbf{2019}, \emph{4}, 971--982\relax
\mciteBstWouldAddEndPuncttrue
\mciteSetBstMidEndSepPunct{\mcitedefaultmidpunct}
{\mcitedefaultendpunct}{\mcitedefaultseppunct}\relax
\EndOfBibitem
\bibitem[Bhalothia \latin{et~al.}(2019)Bhalothia, Lin, Yan, Yang, and Chen]{bhalothia2019effects}
Bhalothia,~D.; Lin,~C.-Y.; Yan,~C.; Yang,~Y.-T.; Chen,~T.-Y. {Effects of Pt metal loading on the atomic restructure and oxygen reduction reaction performance of Pt-cluster decorated Cu@ Pd electrocatalysts}. \emph{Sustainable Energy \& Fuels} \textbf{2019}, \emph{3}, 1668--1681\relax
\mciteBstWouldAddEndPuncttrue
\mciteSetBstMidEndSepPunct{\mcitedefaultmidpunct}
{\mcitedefaultendpunct}{\mcitedefaultseppunct}\relax
\EndOfBibitem
\bibitem[Merchant \latin{et~al.}(2000)Merchant, Wang, Giannuzzi, and Liu]{merchant2000metallurgy}
Merchant,~H.; Wang,~J.; Giannuzzi,~L.; Liu,~Y. {Metallurgy and performance of electrodeposited copper for flexible circuits}. \emph{Circuit World} \textbf{2000}, \emph{26}, 7--14\relax
\mciteBstWouldAddEndPuncttrue
\mciteSetBstMidEndSepPunct{\mcitedefaultmidpunct}
{\mcitedefaultendpunct}{\mcitedefaultseppunct}\relax
\EndOfBibitem
\bibitem[Fotopoulos \latin{et~al.}(2024)Fotopoulos, O'Hern, Shattuck, and Shluger]{fotopoulos2024modeling}
Fotopoulos,~V.; O'Hern,~C.~S.; Shattuck,~M.~D.; Shluger,~A.~L. Modeling the Effects of Varying the {Ti} Concentration on the Mechanical Properties of {Cu--Ti} Alloys. \emph{ACS Omega} \textbf{2024}, \emph{9}, 1482\relax
\mciteBstWouldAddEndPuncttrue
\mciteSetBstMidEndSepPunct{\mcitedefaultmidpunct}
{\mcitedefaultendpunct}{\mcitedefaultseppunct}\relax
\EndOfBibitem
\bibitem[Kleinbichler \latin{et~al.}(2021)Kleinbichler, Kofler, Stabentheiner, Reisinger, Moser, Zechner, Nelhiebel, and Kozeschnik]{kleinbichler2021quantitative}
Kleinbichler,~M.; Kofler,~C.; Stabentheiner,~M.; Reisinger,~M.; Moser,~S.; Zechner,~J.; Nelhiebel,~M.; Kozeschnik,~E. {Quantitative analysis of void initiation in thermo-mechanical fatigue of polycrystalline copper films}. \emph{Microelectronics Reliability} \textbf{2021}, \emph{127}, 114387\relax
\mciteBstWouldAddEndPuncttrue
\mciteSetBstMidEndSepPunct{\mcitedefaultmidpunct}
{\mcitedefaultendpunct}{\mcitedefaultseppunct}\relax
\EndOfBibitem
\bibitem[Moser \latin{et~al.}(2022)Moser, Kleinbichler, Zechner, Reisinger, Nelhiebel, and Cordill]{moser2022fatigue}
Moser,~S.; Kleinbichler,~M.; Zechner,~J.; Reisinger,~M.; Nelhiebel,~M.; Cordill,~M.~J. Fatigue of copper films subjected to high-strain rate thermo-mechanical pulsing. \emph{Microelectronics Reliability} \textbf{2022}, \emph{137}, 114782\relax
\mciteBstWouldAddEndPuncttrue
\mciteSetBstMidEndSepPunct{\mcitedefaultmidpunct}
{\mcitedefaultendpunct}{\mcitedefaultseppunct}\relax
\EndOfBibitem
\bibitem[Arnaud \latin{et~al.}(2003)Arnaud, Berger, and Reimbold]{arnaud2003evidence}
Arnaud,~L.; Berger,~T.; Reimbold,~G. Evidence of grain-boundary versus interface diffusion in electromigration experiments in copper damascene interconnects. \emph{{Journal of Applied Physics}} \textbf{2003}, \emph{93}, 192--204\relax
\mciteBstWouldAddEndPuncttrue
\mciteSetBstMidEndSepPunct{\mcitedefaultmidpunct}
{\mcitedefaultendpunct}{\mcitedefaultseppunct}\relax
\EndOfBibitem
\bibitem[Huang \latin{et~al.}(2018)Huang, Chen, Shen, Zhang, and Rupert]{huang2018uncovering}
Huang,~Z.; Chen,~F.; Shen,~Q.; Zhang,~L.; Rupert,~T.~J. {Uncovering the influence of common nonmetallic impurities on the stability and strength of a {$\Sigma$}5 (310) grain boundary in Cu}. \emph{Acta Materialia} \textbf{2018}, \emph{148}, 110--122\relax
\mciteBstWouldAddEndPuncttrue
\mciteSetBstMidEndSepPunct{\mcitedefaultmidpunct}
{\mcitedefaultendpunct}{\mcitedefaultseppunct}\relax
\EndOfBibitem
\bibitem[Moser \latin{et~al.}(2019)Moser, Zernatto, Kleinbichler, Nelhiebel, Zechner, Cordill, and Pippan]{moser2019novel}
Moser,~S.; Zernatto,~G.; Kleinbichler,~M.; Nelhiebel,~M.; Zechner,~J.; Cordill,~M.~J.; Pippan,~R. {A novel setup for in situ monitoring of thermomechanically cycled thin film metallizations}. \emph{JOM} \textbf{2019}, \emph{71}, 3399--3406\relax
\mciteBstWouldAddEndPuncttrue
\mciteSetBstMidEndSepPunct{\mcitedefaultmidpunct}
{\mcitedefaultendpunct}{\mcitedefaultseppunct}\relax
\EndOfBibitem
\bibitem[Moser \latin{et~al.}(2021)Moser, Kleinbichler, Kubicek, Zechner, and Cordill]{moser2021electropolishing}
Moser,~S.; Kleinbichler,~M.; Kubicek,~S.; Zechner,~J.; Cordill,~M.~J. {Electropolishing---a practical method for accessing voids in metal films for analyses}. \emph{Applied Sciences} \textbf{2021}, \emph{11}, 7009\relax
\mciteBstWouldAddEndPuncttrue
\mciteSetBstMidEndSepPunct{\mcitedefaultmidpunct}
{\mcitedefaultendpunct}{\mcitedefaultseppunct}\relax
\EndOfBibitem
\bibitem[Nakahara(1988)]{nakahara1988microscopic}
Nakahara,~S. Microscopic mechanism of the hydrogen effect on the ductility of electroless copper. \emph{Acta Metallurgica} \textbf{1988}, \emph{36}, 1669--1681\relax
\mciteBstWouldAddEndPuncttrue
\mciteSetBstMidEndSepPunct{\mcitedefaultmidpunct}
{\mcitedefaultendpunct}{\mcitedefaultseppunct}\relax
\EndOfBibitem
\bibitem[Chen \latin{et~al.}(2025)Chen, Huang, Liu, Yen, Niu, Burr, Moore, Mart{\'\i}nez-Pa{\~n}eda, Atrens, and Cairney]{chen2025hydrogen}
Chen,~Y.-S.; Huang,~C.; Liu,~P.-Y.; Yen,~H.-W.; Niu,~R.; Burr,~P.; Moore,~K.~L.; Mart{\'\i}nez-Pa{\~n}eda,~E.; Atrens,~A.; Cairney,~J.~M. Hydrogen trapping and embrittlement in metals--a review. \emph{{International Journal of Hydrogen Energy}} \textbf{2025}, \emph{136}, 789--821\relax
\mciteBstWouldAddEndPuncttrue
\mciteSetBstMidEndSepPunct{\mcitedefaultmidpunct}
{\mcitedefaultendpunct}{\mcitedefaultseppunct}\relax
\EndOfBibitem
\bibitem[Fotopoulos \latin{et~al.}(2023)Fotopoulos, Grau-Crespo, and Shluger]{fotopoulos2023thermodynamic}
Fotopoulos,~V.; Grau-Crespo,~R.; Shluger,~A.~L. {Thermodynamic analysis of the interaction between metal vacancies and hydrogen in bulk Cu}. \emph{Physical Chemistry Chemical Physics} \textbf{2023}, \emph{25}, 9168--9175\relax
\mciteBstWouldAddEndPuncttrue
\mciteSetBstMidEndSepPunct{\mcitedefaultmidpunct}
{\mcitedefaultendpunct}{\mcitedefaultseppunct}\relax
\EndOfBibitem
\bibitem[Alvarez-Falcon \latin{et~al.}(2016)Alvarez-Falcon, Vines, Notario-Estevez, and Illas]{alvarez2016hydrogen}
Alvarez-Falcon,~L.; Vines,~F.; Notario-Estevez,~A.; Illas,~F. On the hydrogen adsorption and dissociation on {Cu} surfaces and nanorows. \emph{Surface Science} \textbf{2016}, \emph{646}, 221--229\relax
\mciteBstWouldAddEndPuncttrue
\mciteSetBstMidEndSepPunct{\mcitedefaultmidpunct}
{\mcitedefaultendpunct}{\mcitedefaultseppunct}\relax
\EndOfBibitem
\bibitem[Mistry \latin{et~al.}(2024)Mistry, Snowden, Darling, and Hodgson]{mistry2024hydroxyl}
Mistry,~K.; Snowden,~H.; Darling,~G.~R.; Hodgson,~A. Hydroxyl on Stepped Copper and its Interaction with Water. \emph{The Journal of Physical Chemistry C} \textbf{2024}, \emph{128}, 13025--13033\relax
\mciteBstWouldAddEndPuncttrue
\mciteSetBstMidEndSepPunct{\mcitedefaultmidpunct}
{\mcitedefaultendpunct}{\mcitedefaultseppunct}\relax
\EndOfBibitem
\bibitem[Jedidi \latin{et~al.}(2015)Jedidi, Rasul, Masih, Cavallo, and Takanabe]{jedidi2015generation}
Jedidi,~A.; Rasul,~S.; Masih,~D.; Cavallo,~L.; Takanabe,~K. Generation of {Cu--In} alloy surfaces from {CuInO}2 as selective catalytic sites for {CO}2 electroreduction. \emph{Journal of Materials Chemistry A} \textbf{2015}, \emph{3}, 19085--19092\relax
\mciteBstWouldAddEndPuncttrue
\mciteSetBstMidEndSepPunct{\mcitedefaultmidpunct}
{\mcitedefaultendpunct}{\mcitedefaultseppunct}\relax
\EndOfBibitem
\bibitem[Amram \latin{et~al.}(2014)Amram, Klinger, Gazit, Gluska, and Rabkin]{amram2014grain}
Amram,~D.; Klinger,~L.; Gazit,~N.; Gluska,~H.; Rabkin,~E. Grain boundary grooving in thin films revisited: the role of interface diffusion. \emph{Acta materialia} \textbf{2014}, \emph{69}, 386--396\relax
\mciteBstWouldAddEndPuncttrue
\mciteSetBstMidEndSepPunct{\mcitedefaultmidpunct}
{\mcitedefaultendpunct}{\mcitedefaultseppunct}\relax
\EndOfBibitem
\bibitem[Bentria \latin{et~al.}(2017)Bentria, N'tsouaglo, Becquart, Bouhali, Mousseau, and El-Mellouhi]{bentria2017role}
Bentria,~E.~T.; N'tsouaglo,~G.~K.; Becquart,~C.~S.; Bouhali,~O.; Mousseau,~N.; El-Mellouhi,~F. The role of emerging grain boundary at iron surface, temperature and hydrogen on metal dusting initiation. \emph{Acta Materialia} \textbf{2017}, \emph{135}, 340--347\relax
\mciteBstWouldAddEndPuncttrue
\mciteSetBstMidEndSepPunct{\mcitedefaultmidpunct}
{\mcitedefaultendpunct}{\mcitedefaultseppunct}\relax
\EndOfBibitem
\bibitem[Momida \latin{et~al.}(2013)Momida, Asari, Nakamura, Tateyama, and Ohno]{momida2013hydrogen}
Momida,~H.; Asari,~Y.; Nakamura,~Y.; Tateyama,~Y.; Ohno,~T. Hydrogen-enhanced vacancy embrittlement of grain boundaries in iron. \emph{Physical Review B---Condensed Matter and Materials Physics} \textbf{2013}, \emph{88}, 144107\relax
\mciteBstWouldAddEndPuncttrue
\mciteSetBstMidEndSepPunct{\mcitedefaultmidpunct}
{\mcitedefaultendpunct}{\mcitedefaultseppunct}\relax
\EndOfBibitem
\bibitem[Du \latin{et~al.}(2020)Du, Geng, Arakawa, Li, and Ogata]{du2020hydrogen}
Du,~J.-P.; Geng,~W.; Arakawa,~K.; Li,~J.; Ogata,~S. {Hydrogen-enhanced vacancy diffusion in metals}. \emph{The Journal of Physical Chemistry Letters} \textbf{2020}, \emph{11}, 7015--7020\relax
\mciteBstWouldAddEndPuncttrue
\mciteSetBstMidEndSepPunct{\mcitedefaultmidpunct}
{\mcitedefaultendpunct}{\mcitedefaultseppunct}\relax
\EndOfBibitem
\bibitem[Sobola and Dallaev(2024)Sobola, and Dallaev]{sobola2024exploring}
Sobola,~D.; Dallaev,~R. Exploring hydrogen embrittlement: mechanisms, consequences, and advances in metal science. \emph{Energies} \textbf{2024}, \emph{17}, 2972\relax
\mciteBstWouldAddEndPuncttrue
\mciteSetBstMidEndSepPunct{\mcitedefaultmidpunct}
{\mcitedefaultendpunct}{\mcitedefaultseppunct}\relax
\EndOfBibitem
\bibitem[Birnbaum and Sofronis(1994)Birnbaum, and Sofronis]{birnbaum1994hydrogen}
Birnbaum,~H.~K.; Sofronis,~P. {Hydrogen-enhanced localized plasticity---a mechanism for hydrogen-related fracture}. \emph{Materials Science and Engineering: A} \textbf{1994}, \emph{176}, 191--202\relax
\mciteBstWouldAddEndPuncttrue
\mciteSetBstMidEndSepPunct{\mcitedefaultmidpunct}
{\mcitedefaultendpunct}{\mcitedefaultseppunct}\relax
\EndOfBibitem
\bibitem[Ganchenkova \latin{et~al.}(2014)Ganchenkova, Yagodzinskyy, Borodin, and H{\"a}nninen]{ganchenkova2014effects}
Ganchenkova,~M.; Yagodzinskyy,~Y.; Borodin,~V.; H{\"a}nninen,~H. {Effects of hydrogen and impurities on void nucleation in copper: simulation point of view}. \emph{Philosophical Magazine} \textbf{2014}, \emph{94}, 3522--3548\relax
\mciteBstWouldAddEndPuncttrue
\mciteSetBstMidEndSepPunct{\mcitedefaultmidpunct}
{\mcitedefaultendpunct}{\mcitedefaultseppunct}\relax
\EndOfBibitem
\bibitem[Lousada and Korzhavyi(2023)Lousada, and Korzhavyi]{lousada2023pathways}
Lousada,~C.~M.; Korzhavyi,~P.~A. Pathways of hydrogen atom diffusion at fcc {Cu}: {$\Sigma$9} and {$\Sigma$5} grain boundaries vs single crystal. \emph{Journal of Materials Science} \textbf{2023}, \emph{58}, 17004--17018\relax
\mciteBstWouldAddEndPuncttrue
\mciteSetBstMidEndSepPunct{\mcitedefaultmidpunct}
{\mcitedefaultendpunct}{\mcitedefaultseppunct}\relax
\EndOfBibitem
\bibitem[Lousada and Korzhavyi(2022)Lousada, and Korzhavyi]{lousada2022single}
Lousada,~C.~M.; Korzhavyi,~P.~A. {Single vacancies at $\Sigma$5, $\Sigma$9 and $\Sigma$11 grain boundaries of copper and the geometrical factors that affect their site preference}. \emph{{Journal of Physics and Chemistry of Solids}} \textbf{2022}, \emph{169}, 110833\relax
\mciteBstWouldAddEndPuncttrue
\mciteSetBstMidEndSepPunct{\mcitedefaultmidpunct}
{\mcitedefaultendpunct}{\mcitedefaultseppunct}\relax
\EndOfBibitem
\bibitem[Huang \latin{et~al.}(2023)Huang, Song, Zhou, Su, Qiao, and Gao]{huang2023hydrogen}
Huang,~C.; Song,~K.; Zhou,~S.; Su,~Y.; Qiao,~L.; Gao,~L. {Hydrogen atom solution and diffusion behaviors at $\Sigma$3 and $\Sigma$5 grain boundaries of Fe, Ni, Cu and Al: A first-principles study}. \emph{{Materials Today Communications}} \textbf{2023}, \emph{37}, 107222\relax
\mciteBstWouldAddEndPuncttrue
\mciteSetBstMidEndSepPunct{\mcitedefaultmidpunct}
{\mcitedefaultendpunct}{\mcitedefaultseppunct}\relax
\EndOfBibitem
\bibitem[Huang \latin{et~al.}(2019)Huang, Chen, Shen, Zhang, and Rupert]{huang2019combined}
Huang,~Z.; Chen,~F.; Shen,~Q.; Zhang,~L.; Rupert,~T.~J. {Combined effects of nonmetallic impurities and planned metallic dopants on grain boundary energy and strength}. \emph{Acta Materialia} \textbf{2019}, \emph{166}, 113--125\relax
\mciteBstWouldAddEndPuncttrue
\mciteSetBstMidEndSepPunct{\mcitedefaultmidpunct}
{\mcitedefaultendpunct}{\mcitedefaultseppunct}\relax
\EndOfBibitem
\bibitem[Bodlos \latin{et~al.}(2023)Bodlos, Scheiber, Spitaler, and Romaner]{bodlos2023modification}
Bodlos,~R.; Scheiber,~D.; Spitaler,~J.; Romaner,~L. {Modification of the Cu/W interface cohesion by segregation}. \emph{Metals} \textbf{2023}, \emph{13}, 346\relax
\mciteBstWouldAddEndPuncttrue
\mciteSetBstMidEndSepPunct{\mcitedefaultmidpunct}
{\mcitedefaultendpunct}{\mcitedefaultseppunct}\relax
\EndOfBibitem
\bibitem[Fotopoulos \latin{et~al.}(2024)Fotopoulos, Strand, Petersmann, and Shluger]{fotopoulos2024first}
Fotopoulos,~V.; Strand,~J.; Petersmann,~M.; Shluger,~A.~L. First principles study on the segregation of metallic solutes and non-metallic impurities in {Cu} grain boundary. TMS Annual Meeting \& Exhibition. 2024; pp 989--999\relax
\mciteBstWouldAddEndPuncttrue
\mciteSetBstMidEndSepPunct{\mcitedefaultmidpunct}
{\mcitedefaultendpunct}{\mcitedefaultseppunct}\relax
\EndOfBibitem
\bibitem[Lousada and Korzhavyi(2020)Lousada, and Korzhavyi]{lousada2020hydrogen}
Lousada,~C.~M.; Korzhavyi,~P.~A. {Hydrogen sorption capacity of crystal lattice defects and low Miller index surfaces of copper}. \emph{Journal of Materials Science} \textbf{2020}, \emph{55}, 6623--6636\relax
\mciteBstWouldAddEndPuncttrue
\mciteSetBstMidEndSepPunct{\mcitedefaultmidpunct}
{\mcitedefaultendpunct}{\mcitedefaultseppunct}\relax
\EndOfBibitem
\bibitem[Kresse and Hafner(1993)Kresse, and Hafner]{kresse1993ab}
Kresse,~G.; Hafner,~J. {Ab initio molecular dynamics for open-shell transition metals}. \emph{Physical Review B} \textbf{1993}, \emph{48}, 13115\relax
\mciteBstWouldAddEndPuncttrue
\mciteSetBstMidEndSepPunct{\mcitedefaultmidpunct}
{\mcitedefaultendpunct}{\mcitedefaultseppunct}\relax
\EndOfBibitem
\bibitem[Kresse and Furthm{\"u}ller(1996)Kresse, and Furthm{\"u}ller]{kresse1996efficient}
Kresse,~G.; Furthm{\"u}ller,~J. {Efficient iterative schemes for ab initio total-energy calculations using a plane-wave basis set}. \emph{Physical Review B} \textbf{1996}, \emph{54}, 11169\relax
\mciteBstWouldAddEndPuncttrue
\mciteSetBstMidEndSepPunct{\mcitedefaultmidpunct}
{\mcitedefaultendpunct}{\mcitedefaultseppunct}\relax
\EndOfBibitem
\bibitem[Kresse and Furthm{\"u}ller(1996)Kresse, and Furthm{\"u}ller]{kresse1996efficiency}
Kresse,~G.; Furthm{\"u}ller,~J. {Efficiency of ab-initio total energy calculations for metals and semiconductors using a plane-wave basis set}. \emph{Computational Materials Science} \textbf{1996}, \emph{6}, 15--50\relax
\mciteBstWouldAddEndPuncttrue
\mciteSetBstMidEndSepPunct{\mcitedefaultmidpunct}
{\mcitedefaultendpunct}{\mcitedefaultseppunct}\relax
\EndOfBibitem
\bibitem[Perdew \latin{et~al.}(1996)Perdew, Burke, and Ernzerhof]{perdew1996generalized}
Perdew,~J.~P.; Burke,~K.; Ernzerhof,~M. {Generalized gradient approximation made simple}. \emph{Physical Review Letters} \textbf{1996}, \emph{77}, 3865\relax
\mciteBstWouldAddEndPuncttrue
\mciteSetBstMidEndSepPunct{\mcitedefaultmidpunct}
{\mcitedefaultendpunct}{\mcitedefaultseppunct}\relax
\EndOfBibitem
\bibitem[Davidson(1975)]{er1975iterativecalculationof}
Davidson,~E.~R. {The iterative calculation of a few of the lowest eigenvalues and corresponding eigenvectors of large real-symmetric matrices}. \emph{Journal of Computational Physics} \textbf{1975}, \emph{17}, 87--94\relax
\mciteBstWouldAddEndPuncttrue
\mciteSetBstMidEndSepPunct{\mcitedefaultmidpunct}
{\mcitedefaultendpunct}{\mcitedefaultseppunct}\relax
\EndOfBibitem
\bibitem[Pulay(1980)]{pulay1980convergence}
Pulay,~P. {Convergence acceleration of iterative sequences. The case of SCF iteration}. \emph{Chemical Physics Letters} \textbf{1980}, \emph{73}, 393--398\relax
\mciteBstWouldAddEndPuncttrue
\mciteSetBstMidEndSepPunct{\mcitedefaultmidpunct}
{\mcitedefaultendpunct}{\mcitedefaultseppunct}\relax
\EndOfBibitem
\bibitem[Wood and Zunger(1985)Wood, and Zunger]{wood1985new}
Wood,~D.; Zunger,~A. {A new method for diagonalising large matrices}. \emph{Journal of Physics A: Mathematical and General} \textbf{1985}, \emph{18}, 1343\relax
\mciteBstWouldAddEndPuncttrue
\mciteSetBstMidEndSepPunct{\mcitedefaultmidpunct}
{\mcitedefaultendpunct}{\mcitedefaultseppunct}\relax
\EndOfBibitem
\bibitem[Thompson \latin{et~al.}(2022)Thompson, Aktulga, Berger, Bolintineanu, Brown, Crozier, in't Veld, Kohlmeyer, Moore, Nguyen, Shan, Stevens, Tranchida, Trott, and Plimpton]{thompson2022lammps}
Thompson,~A.~P.; Aktulga,~H.~M.; Berger,~R.; Bolintineanu,~D.~S.; Brown,~W.~M.; Crozier,~P.~S.; in't Veld,~P.~J.; Kohlmeyer,~A.; Moore,~S.~G.; Nguyen,~T.~D.; Shan,~R.; Stevens,~M.~J.; Tranchida,~J.; Trott,~C.; Plimpton,~S.~J. {LAMMPS-a flexible simulation tool for particle-based materials modeling at the atomic, meso, and continuum scales}. \emph{Computer Physics Communications} \textbf{2022}, \emph{271}, 108171\relax
\mciteBstWouldAddEndPuncttrue
\mciteSetBstMidEndSepPunct{\mcitedefaultmidpunct}
{\mcitedefaultendpunct}{\mcitedefaultseppunct}\relax
\EndOfBibitem
\bibitem[Zhou \latin{et~al.}(2015)Zhou, Ward, Foster, and Zimmerman]{zhou2015analytical}
Zhou,~X.; Ward,~D.; Foster,~M.; Zimmerman,~J. {An analytical bond-order potential for the copper--hydrogen binary system}. \emph{Journal of Materials Science} \textbf{2015}, \emph{50}, 2859--2875\relax
\mciteBstWouldAddEndPuncttrue
\mciteSetBstMidEndSepPunct{\mcitedefaultmidpunct}
{\mcitedefaultendpunct}{\mcitedefaultseppunct}\relax
\EndOfBibitem
\bibitem[Zhou \latin{et~al.}(2016)Zhou, Ward, and Foster]{zhou2016analytical}
Zhou,~X.; Ward,~D.; Foster,~M. {An analytical bond-order potential for the aluminum copper binary system}. \emph{Journal of Alloys and Compounds} \textbf{2016}, \emph{680}, 752--767\relax
\mciteBstWouldAddEndPuncttrue
\mciteSetBstMidEndSepPunct{\mcitedefaultmidpunct}
{\mcitedefaultendpunct}{\mcitedefaultseppunct}\relax
\EndOfBibitem
\bibitem[Henkelman and J{\'o}nsson(2000)Henkelman, and J{\'o}nsson]{henkelman2000improved}
Henkelman,~G.; J{\'o}nsson,~H. {Improved tangent estimate in the nudged elastic band method for finding minimum energy paths and saddle points}. \emph{The Journal of Chemical Physics} \textbf{2000}, \emph{113}, 9978--9985\relax
\mciteBstWouldAddEndPuncttrue
\mciteSetBstMidEndSepPunct{\mcitedefaultmidpunct}
{\mcitedefaultendpunct}{\mcitedefaultseppunct}\relax
\EndOfBibitem
\bibitem[Henkelman \latin{et~al.}(2000)Henkelman, Uberuaga, and J{\'o}nsson]{henkelman2000climbing}
Henkelman,~G.; Uberuaga,~B.~P.; J{\'o}nsson,~H. {A climbing image nudged elastic band method for finding saddle points and minimum energy paths}. \emph{The Journal of Chemical Physics} \textbf{2000}, \emph{113}, 9901--9904\relax
\mciteBstWouldAddEndPuncttrue
\mciteSetBstMidEndSepPunct{\mcitedefaultmidpunct}
{\mcitedefaultendpunct}{\mcitedefaultseppunct}\relax
\EndOfBibitem
\bibitem[Maras \latin{et~al.}(2016)Maras, Trushin, Stukowski, Ala-Nissila, and Jonsson]{maras2016global}
Maras,~E.; Trushin,~O.; Stukowski,~A.; Ala-Nissila,~T.; Jonsson,~H. {Global transition path search for dislocation formation in Ge on Si (001)}. \emph{Computer Physics Communications} \textbf{2016}, \emph{205}, 13--21\relax
\mciteBstWouldAddEndPuncttrue
\mciteSetBstMidEndSepPunct{\mcitedefaultmidpunct}
{\mcitedefaultendpunct}{\mcitedefaultseppunct}\relax
\EndOfBibitem
\bibitem[Nakano(2008)]{nakano2008space}
Nakano,~A. {A space--time-ensemble parallel nudged elastic band algorithm for molecular kinetics simulation}. \emph{Computer Physics Communications} \textbf{2008}, \emph{178}, 280--289\relax
\mciteBstWouldAddEndPuncttrue
\mciteSetBstMidEndSepPunct{\mcitedefaultmidpunct}
{\mcitedefaultendpunct}{\mcitedefaultseppunct}\relax
\EndOfBibitem
\bibitem[Ding \latin{et~al.}(2024)Ding, Yu, Lin, Ortiz, Xiao, He, and Zhang]{ding2024hydrogen}
Ding,~Y.; Yu,~H.; Lin,~M.; Ortiz,~M.; Xiao,~S.; He,~J.; Zhang,~Z. Hydrogen trapping and diffusion in polycrystalline nickel: The spectrum of grain boundary segregation. \emph{Journal of Materials Science \& Technology} \textbf{2024}, \emph{173}, 225--236\relax
\mciteBstWouldAddEndPuncttrue
\mciteSetBstMidEndSepPunct{\mcitedefaultmidpunct}
{\mcitedefaultendpunct}{\mcitedefaultseppunct}\relax
\EndOfBibitem
\bibitem[Grimme(2006)]{grimme2006semiempirical}
Grimme,~S. Semiempirical {GGA}-type density functional constructed with a long-range dispersion correction. \emph{Journal of Computational Chemistry} \textbf{2006}, \emph{27}, 1787--1799\relax
\mciteBstWouldAddEndPuncttrue
\mciteSetBstMidEndSepPunct{\mcitedefaultmidpunct}
{\mcitedefaultendpunct}{\mcitedefaultseppunct}\relax
\EndOfBibitem
\bibitem[Fotopoulos \latin{et~al.}(2023)Fotopoulos, Mora-Fonz, Kleinbichler, Bodlos, Kozeschnik, Romaner, and Shluger]{fotopoulos2023structure}
Fotopoulos,~V.; Mora-Fonz,~D.; Kleinbichler,~M.; Bodlos,~R.; Kozeschnik,~E.; Romaner,~L.; Shluger,~A.~L. {{Structure and Migration Mechanisms of Small Vacancy Clusters in Cu: A Combined EAM and DFT Study}}. \emph{Nanomaterials} \textbf{2023}, \emph{13}, 1464\relax
\mciteBstWouldAddEndPuncttrue
\mciteSetBstMidEndSepPunct{\mcitedefaultmidpunct}
{\mcitedefaultendpunct}{\mcitedefaultseppunct}\relax
\EndOfBibitem
\bibitem[Katz \latin{et~al.}(1971)Katz, Guinan, and Borg]{katz1971diffusion}
Katz,~L.; Guinan,~M.; Borg,~R. Diffusion of {H}2, {D}2, and {T}2 in single-crystal {Ni} and {Cu}. \emph{Physical Review B} \textbf{1971}, \emph{4}, 330\relax
\mciteBstWouldAddEndPuncttrue
\mciteSetBstMidEndSepPunct{\mcitedefaultmidpunct}
{\mcitedefaultendpunct}{\mcitedefaultseppunct}\relax
\EndOfBibitem
\bibitem[Zhou \latin{et~al.}(2013)Zhou, Zhang, and Ou]{zhou2013dissolution}
Zhou,~H.-B.; Zhang,~Y.; Ou,~X. Dissolution and diffusion behaviors of hydrogen in copper: A first-principles investigation. \emph{Computational Materials Science} \textbf{2013}, \emph{79}, 923--928\relax
\mciteBstWouldAddEndPuncttrue
\mciteSetBstMidEndSepPunct{\mcitedefaultmidpunct}
{\mcitedefaultendpunct}{\mcitedefaultseppunct}\relax
\EndOfBibitem
\bibitem[Williams and Galindo-Nava(2023)Williams, and Galindo-Nava]{williams2023accelerating}
Williams,~C.; Galindo-Nava,~E. Accelerating off-lattice kinetic {Monte Carlo} simulations to predict hydrogen vacancy-cluster interactions in $\alpha$-{Fe}. \emph{Acta Materialia} \textbf{2023}, \emph{242}, 118452\relax
\mciteBstWouldAddEndPuncttrue
\mciteSetBstMidEndSepPunct{\mcitedefaultmidpunct}
{\mcitedefaultendpunct}{\mcitedefaultseppunct}\relax
\EndOfBibitem
\bibitem[Steffen and Alibakhshi(2024)Steffen, and Alibakhshi]{steffen2024hydrogen}
Steffen,~J.; Alibakhshi,~A. Hydrogen diffusion on {Ni} (100): A combined machine-learning, ring polymer molecular dynamics, and kinetic {Monte Carlo} study. \emph{The Journal of Chemical Physics} \textbf{2024}, \emph{161}\relax
\mciteBstWouldAddEndPuncttrue
\mciteSetBstMidEndSepPunct{\mcitedefaultmidpunct}
{\mcitedefaultendpunct}{\mcitedefaultseppunct}\relax
\EndOfBibitem
\bibitem[Navidtehrani \latin{et~al.}(2025)Navidtehrani, Beteg{\'o}n, and Mart{\'\i}nez-Pa{\~n}eda]{navidtehrani2025generalised}
Navidtehrani,~Y.; Beteg{\'o}n,~C.; Mart{\'\i}nez-Pa{\~n}eda,~E. A generalised framework for phase field-based modelling of coupled problems: Application to thermo-mechanical fracture, hydraulic fracture, hydrogen embrittlement and corrosion. \emph{{Engineering Fracture Mechanics}} \textbf{2025}, 111363\relax
\mciteBstWouldAddEndPuncttrue
\mciteSetBstMidEndSepPunct{\mcitedefaultmidpunct}
{\mcitedefaultendpunct}{\mcitedefaultseppunct}\relax
\EndOfBibitem
\bibitem[Kristensen \latin{et~al.}(2020)Kristensen, Niordson, and Mart{\'\i}nez-Pa{\~n}eda]{kristensen2020applications}
Kristensen,~P.~K.; Niordson,~C.~F.; Mart{\'\i}nez-Pa{\~n}eda,~E. Applications of phase field fracture in modelling hydrogen assisted failures. \emph{Theoretical and Applied Fracture Mechanics} \textbf{2020}, \emph{110}, 102837\relax
\mciteBstWouldAddEndPuncttrue
\mciteSetBstMidEndSepPunct{\mcitedefaultmidpunct}
{\mcitedefaultendpunct}{\mcitedefaultseppunct}\relax
\EndOfBibitem
\end{mcitethebibliography}

\end{document}